\documentclass[sigconf]{acmart}

\usepackage{amsmath,amssymb,amsfonts}
\usepackage{graphicx}
\usepackage{xcolor}
\usepackage{subcaption}
\usepackage{hyperref}

\AtBeginDocument{%
  \providecommand\BibTeX{{%
    \normalfont B\kern-0.5em{\scshape i\kern-0.25em b}\kern-0.8em\TeX}}}

\copyrightyear{2021}
\acmYear{2021}
\setcopyright{licensedothergov}\acmConference[KDD '21]{Proceedings of the 27th ACM SIGKDD Conference on Knowledge Discovery and Data Mining}{August 14--18, 2021}{Virtual Event, Singapore}
\acmBooktitle{Proceedings of the 27th ACM SIGKDD Conference on Knowledge Discovery and Data Mining (KDD '21), August 14--18, 2021, Virtual Event, Singapore}
\acmPrice{15.00}
\acmDOI{10.1145/3447548.3467366}
\acmISBN{978-1-4503-8332-5/21/08}

\begin{document}
\fancyhead{}

\title{Redescription Model Mining}

\author{Felix I. Stamm}
\email{felix.stamm@cssh.rwth-aachen.de}
\affiliation{%
  \institution{RWTH Aachen University}
  \country{Germany}
}

\author{Martin Becker}
\email{becker@informatik.uni-wuerzburg.de}
\affiliation{%
  \institution{University of Würzburg}
  \country{Germany}
}

\author{Markus Strohmaier}
\email{markus.strohmaier@cssh.rwth-aachen.de}
\affiliation{%
  \institution{RWTH Aachen University \& GESIS}
  \country{Germany}
}

\author{Florian Lemmerich}
\email{florian.lemmerich@uni-passau.de}
\affiliation{%
  \institution{University of Passau}
  \country{Germany}
}

\renewcommand{\shortauthors}{F. Stamm, M. Becker, M. Strohmaier, and F. Lemmerich}




\begin{abstract}
This paper introduces Redescription Model Mining, a novel approach to identify interpretable patterns across two datasets that share only a subset of attributes and have no common instances.
In particular, Redescription Model Mining aims to find pairs of describable data subsets -- one for each dataset -- that induce similar exceptional models with respect to a prespecified model class.
To achieve this, we combine two previously separate research areas: Exceptional Model Mining and Redescription Mining.
For this new problem setting, we develop interestingness measures to select promising patterns, propose efficient algorithms, and demonstrate their potential on synthetic and real-world data. 
Uncovered patterns can hint at common underlying phenomena 
that manifest themselves across datasets, enabling the discovery of possible associations between (combinations of) attributes that do not appear in the same dataset.
\end{abstract}

\maketitle

 \section{Introduction}
 \label{ref:introduction}

In many domains, extensive related datasets exist that are not directly compatible with each other, either because they were collected independently of each other or because data collection evolved over time.
Thus, while they share some common key properties of the domain, they differ with respect to other attributes.
With current state-of-the-art techniques, it is difficult to associate attributes across datasets if they do not share instances.
Yet, unveiling associations across datasets carries the potential to combine information from larger amounts of data as well as complementary data sources.

\begin{figure}[t]
		\includegraphics[width=0.98\columnwidth]{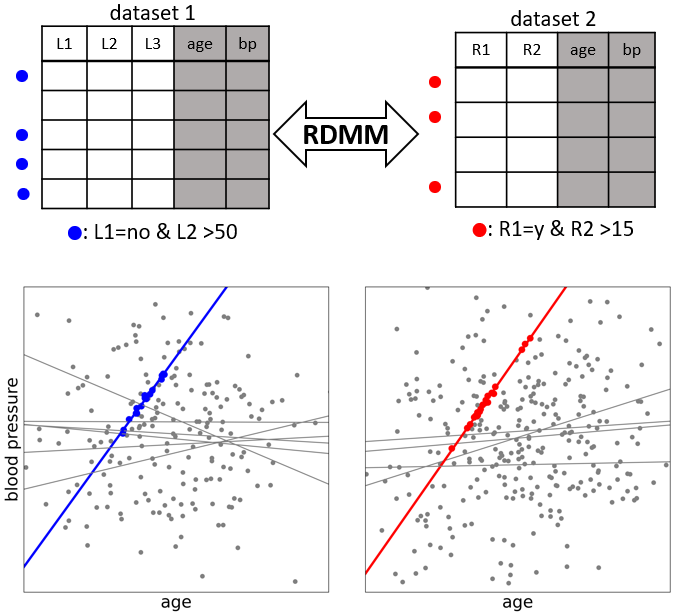}	\label{fig:rL}
		\caption{Redescription Model Mining illustration. We show two toy datasets with different attributes and instances, but shared model attributes, i.e., age and blood pressure (bp).
		In the entire datasets as well as in many subgroups, the correlation between these two model attributes is low (indicated by respective linear regression lines in grey). However, we can find subgroups with different descriptions in each dataset that exhibit an exceptionally strong correlation between age and blood pressure (shown in blue/red) hinting at a common underlying process. Redescription Model Mining aims to automatically find such complementary pairs of descriptions in large sets of candidates that uncover hidden phenomena captured across distinct datasets.
		}
\end{figure}

\noindent\textbf{Objectives. } 
To address this issue, the goal of this work is to enable the automatic discovery of pairs of describable subsets across different datasets with similar but exceptional model characteristics. 

\noindent\textbf{Approach. } To this end, we present a novel approach we call \emph{Redescription Model Mining} (RDMM) that combines two previously distinct pattern mining areas: Exceptional Model Mining ~\cite{leman2008,duivesteijn2010} and Redescription Mining~\cite{galbrunRDM}. Exceptional Model Mining aims to find interpretable descriptions of subsets that induce extraordinary parameters for a given class of models, e.g., a particularly extreme slope in a regression model. 
On the other hand, Redescription Mining identifies pairs of distinct interpretable descriptions that describe very similar subsets of instances in a dataset, but relies on the existence of a common set of data instances.
To find potential associations across datasets, we join these two approaches: 
In \emph{Redescription Model Mining} we aim to find pairs of distinct descriptions \emph{across a pair of datasets} that induce similar exceptional models with regard to model attributes shared by both datasets.
Finding those patterns, i.e., pairs of descriptions, can reveal an underlying phenomenon that manifests itself differently in different data sources.

\noindent\textbf{Illustration. }As an example, we may find in one dataset that people who work more than 50 hours per week and have no spouse have significantly higher blood pressure as they get older.
That is, for this subgroup of people the regression line fitted to the patients blood pressure vs. age has an unusually steep slope.
In another dataset, we may find that people who received financial aid for more than 15 years show similarly abnormal slope.
This may point towards a common, previously hidden, underlying phenomenon, e.g., that both groups have a high percentage of smokers, which in turn leads to the observed increase of blood pressure over time. This hypothesis could then be validated in a subsequent study.

We visualize the concept of identifying corresponding exceptional models in different datasets in Figure~1, which shows two datasets that share the two model attributes `age' and `bp' (blood pressure) but do not contain the same instances and other attributes.
In this dataset, we can fit regression models for many different subgroups (i.e., describable subsets such as $male \wedge diabetes$), indicated by grey regression lines.
While most of these models do not have any notable counterparts in the respectively other dataset, we observe two matching significant outliers, marked in blue/red color.
With Redescription Model Mining, we aim to uncover such exceptional correspondences of subgroups on a model level.
By doing so, we can redescribe a phenomenon (model) discovered in one dataset with another description using attributes from the other dataset.
We argue that such hypotheses are often promising candidates for subsequent analysis as they can point to associations between attributes that cannot be found in either dataset alone.

\noindent\textbf{Contributions. } 
In this paper, we propose a novel problem setting that we call Redescription Model Mining.
We provide a formal problem definition, propose selection criteria that capture interesting patterns, discuss algorithmic solutions for mining such patterns, and demonstrate its applicability on synthetic as well as real-world data.
To the authors' knowledge this constitutes the first pattern mining approach that is able to unveil potential associations between features that are contained in different datasets that do not share any instances.
Thus, Redescription Model Mining paves a way to unveil previously undetectable phenomena and their potential causes by leveraging previously incompatible, complementary data sources.

\section{Background}
Our approach builds upon two established pattern mining techniques: Exceptional Model Mining and Redescription Mining. This section summarizes both methods, introduces formal notations, and recapitulates further related work. 
\vspace{-5mm}
\subsection{Exceptional Model Mining}
Exceptional Model Mining (EMM)~\cite{leman2008,duivesteijn2010} represents a generalization of Subgroup Discovery~\cite{Herrera2010}. 
Overall, the goal of Exceptional Model Mining is to find descriptions that identify subsets of the data exhibiting an interesting parameter distribution with respect to a predefined target model class.
For formal notation, we define a dataset $\Omega$ as a set of tuples (instances) $x$ 
which in turn represent values for a set of attributes $\mathcal A$.
EMM distinguishes between two types of attributes: 
describing attributes $\mathcal{A}_D \subset \mathcal{A}$ and 
model attributes $\mathcal{A}_M \subset \mathcal{A}$. 
Descriptive attributes are used to form descriptions of data subsets, model attributes are used in models of a specific class to fit model parameters.
A feature (selector) $f_A$ of an attribute $A \in \mathcal {A}$ is a binary function $f_A \colon \text{values}(A) \to \{0,1\}$. Each feature has an associated set of instances for a dataset $\Omega$ called the support set (also cover) $\text{supp}_\Omega(f):= \{x \in \Omega| f(x) = 1\}$.
In pattern mining, it is common to combine features with logical expressions like `and' or `or' to form a pattern or subgroup description $D$. 
While our approach works in principle with any \emph{pattern description language}, we focus in our experiments on a commonly applied one:
We build features as attribute/value pairs for nominal attributes and intervals for numeric attributes, and form descriptions as conjunctions of features.
Hence, an example description of a subgroup could be:
$\mathit{Age >  50 \wedge country=\text{'US'}}$.
Due to the combinatorial explosion for conjunctions, an exponentially large number of subgroups can be formed even from comparatively few features.

From this large set of candidates, Exceptional Model Mining aims to identify ``interesting'' subgroups.
Traditionally, interestingness implies that the model parameters induced by the model fitted to the subgroup are significantly different from the model parameters derived from all dataset instances.
See~\cite{EMM-2016} for a discussion on the advantages and disadvantages of comparing subgroups against the entire data vs the complement.
Formally, such interesting parameter contrasts are captured by an exceptionality measure $\text{ex}$ (sometimes called quality function or interestingness measure).

For example, consider a study on a student dataset that investigates the correlation (model class) of the two model attributes `exam preparation time' and `final score'.
Other attributes such as age and gender constitute the descriptive attributes.
Fitting the correlation model to a subgroup $D$ gives a value for the correlation, which represents the model parameter $\rho_D$ of that subgroup.
We then design our exceptionality measure to designate subgroups as ``interesting'' if the difference of the subgroup's parameter ($\rho_D$) and the dataset's parameter ($\rho_\Omega$) is large.
A finding of Exceptional Model Mining for this setup can be:
\emph{``While overall there is a positive correlation between the training time and the performance score ($\rho_\Omega=0.35$), the subgroup of males that are older than 50 years exhibits a negative correlation ($\rho_D=-0.1$)''}.

For Exceptional Model Mining, a wide range of different model classes have been studied in literature including correlation models~\cite{downar2017exceptionally}, Bayesian Networks~\cite{duivesteijn2010}, Markov Chain models~\cite{lemmerich2016mining}, agreement models~\cite{belfodil2019deviant} and regression models~\cite{EMM-slopes}.

\subsection{Redescription Mining}
Redescription Mining is another pattern mining technique that aims to uncover pairs of descriptions, which are interesting in the sense that they cover almost the same instance sets.
The attributes that are used in each of the descriptions are selected such that they originate from different views, i.e., different subsets of attributes.
Formally, we assume one dataset $\Omega$ and divide the attributes $\mathcal{A}$ into two sets of describing attributes $\mathcal{A}_{D, L/R}$ with $\mathcal{A}_{D, L} \cup \mathcal{A}_{D, R} = \mathcal A$ and usually also $\mathcal{A}_{D, L} \cap \mathcal{A}_{D, R} = \varnothing$.
Thus, for Redescription Mining, these views give information on different aspects
of the same instances.
By contrast, for \emph{Redescription Model Mining} we will use the same subscripts $L,R$ to refer to distinct datasets.

As an example, consider a dataset where the instances are regions on earth, while one view contains climate related attributes and the other view contains attributes indicating the presence/absence of animals \cite{RDM-ReReMi}.
The goal is to find regional links between climate and animal populations, represented by corresponding pairs of descriptions $(D_L, D_R)$ such that the Jaccard index  $\frac{|\text{supp}(D_L)\cap\text{supp}(D_R)|}{|\text{supp}(D_L)\cup\text{supp}(D_R)|}$ is high.
For a high Jaccard coefficient, discovered correspondences
can be seen as approximate equivalence relationships\cite{RDM-ClusRM} in this dataset.
While Redescription Mining is a powerful approach to find such patterns and relationships, the underlying requirements for the dataset(s) are strong and can be hard to achieve in practice.

\subsection{Further Related Work}
Connecting multiple data tables with each other has been proposed in the context of pattern mining as multi-relational data mining~\cite{dvzeroski2003multi}. However, such approaches join those data tables on instance identities, while we do not assume common instances in our approach.

Other approaches that combine multiple datasets can be found in the field of transfer learning~\cite{trans}.
Transfer learning aims at exploiting the knowledge that learners have accumulated in one dataset for a specific, often predictive task to apply in another dataset. In contrast to our work, they do not aim at extracting explicit, interpretable knowledge across datasets.
In a similar direction, pretraining of embeddings has  gained high popularity for text \cite{devlin2018bert} and image data\cite{yanai2015food}. 
These methods do establish patterns from a large corpus to apply to another dataset, but do not focus on obtaining interpretable relationships between these datasets.

\vspace{5mm}
\section{Redescription Model Mining}
\label{sec:main}
This section presents \emph{Redescription Model Mining}, a novel approach for mining patterns across two datasets. We outline the main idea, introduce a formal problem definition, discuss a framework for suitable interestingness measures and present mining algorithms.

\subsection{Approach}

We aim to uncover patterns across two different datasets that do not share any instances and only have the model attributes in common. In particular, we unveil potential relationships between pairs of descriptions $D_L$ and $D_R$ ---one for each dataset--- that induce similar  "characteristic'' (exceptional) models with respect to the model attributes and chosen model class.
That is, if we fit a model (from a prechosen model class) with the subgroup instances covered by $D_L$ in the left dataset, and fit a model (of the same class) with the instances covered by $D_R$ in the right dataset, an \emph{interesting} pair of models should be \emph{unusually similar}.
Thus, as discussed in more detail below, we assume potentially interesting pairs of patterns to 
(i) cover a substantial number of instances, 
(ii) induce exceptional models in the individual datasets,
(iii) and induce similar models in the two datasets.
We will quantify those properties in an aggregated way by an interestingness measure that assigns a score to each candidate pair of descriptions.
To identify the most interesting (i.e., highest scoring) patterns, we will search in the space of all \emph{pairs of descriptions} with one description referring to one dataset, and the second description referring to the other dataset.

The patterns, i.e., pairs of descriptions, identified by Redescription Model Mining establish a correspondence not on the instance level but on the subgroup level.
Thus, while such patterns cannot guarantee a dependency between the found patterns, they can provide promising starting points for further investigations on the pattern.
From an Exceptional Model Mining perspective, we can view Redescription Model Mining as a parallel search for exceptional models in two datasets with an extended interesting measure that additionally assesses the similarity between the models induced in both datasets.
From a Redescription Mining perspective, Redescription Model Mining replaces the association criterion from  ``these subgroups cover almost the same instances'' with ``these subgroups show an almost identical unusual phenomenon in the model''.

\subsection{Formal Problem Definition}\label{sec:formal}

We define the \emph{top-$k$ Redescription Model Mining} task as:
Given two datasets $\Omega_L, \Omega_R$, corresponding description languages $\mathcal{P}_L, \mathcal{P}_R$, an interestingness measure $\phi:\mathcal{P}_L \times \mathcal{P}_R \to \mathbb{R}$, and a positive integer $k$, 
the goal of top-$k$ Redescription Model Mining is to find the ordered result list $S=((D^L_1, D^R_1), ..., (D^L_k, D^R_k))$ (with $|S|=k$) of description pairs in $\mathcal{P}_L \times \mathcal{P}_R$ such that \\
(i) $\forall i,j~\text{with}~i < j \leq k\colon\quad \phi(D^L_i, D^R_i) \geq \phi(D^L_j, D^R_j)$ (i.e., the description pairs are ordered by interestingness scores) and\\ (ii)
$\forall {P \in (\mathcal{P}_L \times \mathcal{P}_R)  \setminus S}\colon \phi(P) \leq \phi(D^L_k, D^R_k)$ (i.e., pairs in the list have a higher score than any pair not part of the top-k).

For our approach, we assume that the datasets $\Omega_L$ and $\Omega_R$ have a shared set of model attributes $\mathcal{A}_{M}$ but different sets of describing attributes $\mathcal{A}_{D, L}$ and $\mathcal{A}_{D, R}$, which can, but do not have to, overlap.
Both datasets are additionally presumed to have no (or insignificantly few) instances in common, i.e. $\Omega_L \cap \Omega_R = \emptyset$.
As in traditional Exceptional Model Mining, we can in principle employ any pattern description languages $\mathcal{P}_{L/R}$ but will focus on conjunctions of attribute/value pairs and intervals for our experiments.
The choice of an interestingness measure involves choosing a specific model class and is discussed in detail below.

Additionally, we can require all description pairs in the result list to satisfy specific constraints. 
Typical constraints include that the same attribute cannot be used in both descriptions of a pair, or that the description complexity (number of conjunctive clauses in a description) is limited.

\subsection{Interestingness Measures}
\label{sec:interestingness}
The interestingness measure for a Redescription Model Mining task defines which patterns are reported to the data analyst. 
For traditional pattern and rule mining tasks, a large variety of measures have been proposed, see~\cite{geng2006interestingness} for an overview.

\noindent\textbf{Structure of interestingness measures. }
For Exceptional Model Mining, interestingness measures often involve two main components that are weighed against each other: The subgroup size (number of instances covered by the subgroup) and a measure of exceptionality that quantifies how unusual the target model parameters induced by the subgroup are compared to what is derived from the general population (or, alternatively, the subgroup complement).
For Redescription Model Mining, we extend this notion and distinguish between three components that should be involved in the scoring of a candidate pattern:
First, the associated models derived from the instances covered by the two patterns in the two datasets should be \emph{similar} as the main intuition of Redescription Model Mining is to find similar underlying phenomena.
Second, the induced models should be \emph{exceptional}
with respect to the general population model in the respective dataset. 
This facilitates that correspondences across datasets are less likely due to chance and thus reduces the multiple comparison problem.
Finally, the patterns for both datasets should be \emph{large}, that is, should cover many instances. This makes the respective patterns more relevant and 
lowers the chance of random similarities and exceptionalities in the model parameters.

To facilitate the use of these components, we propose the following general structure of interestingness measures for Redescription Model Mining: 
\begin{align*}
\phi(D_L, D_R) = \phi_{\text{size}}^{\alpha_s}(D_L, D_R) \cdot \phi ^{\alpha_e} _{\text{ex}}(D_L, D_R) \cdot \phi_{\text{sim}}(D_L, D_R),
\end{align*}
where $\phi_{\text{size}}$ is a measure of the subgroups' sizes, $\phi_{\text{ex}}$ is a measure that quantifies the exceptionality of the two models derived from the subgroups compared to the models derived from respective overall datasets, and $\phi_{\text{sim}}$ is a measure of similarity between these two models.
The exponents $\alpha_s$ and $\alpha_e$ are parameters chosen by the analyst that allow for emphasizing or de-emphasizing the size or the exceptionality term.
While alternative interestingness measure definitions are also viable, this structure provides a flexible framework that decomposes the overall challenge of finding suitable interestingness measures into more tractable subproblems while also enabling interactive and iterative exploration of different interestingness measures via the choice of different $\alpha$ values.

\noindent\textbf{Size functions. }
The size function $\phi_{\text{size}}$ can be implemented in various straightforward ways: 
First, for each of the descriptions in the pattern, we can, for example, calculate an individual size function based on the (absolute or relative) number of instances covered by a subgroup, or the entropy of the split into subgroup and complement in the respective dataset.
Then, to obtain a size function for the full pattern (including both descriptions), we apply an aggregation function $agg$ on the two individual sizes for the left and right dataset.
To ensure reasonable subgroup sizes in both datasets, we propose to use the minimum of the individual size functions instead of a mean-based aggregation.

\noindent\textbf{Model exceptionality. }
For constructing measures of model exceptionality, different options have been discussed and applied in Exceptional Model Mining literature. It has been proposed to use direct difference measures for the model parameters or structure~\cite{EMM-2016}, to apply bootstrap sampling over those differences~\cite{lemmerich2016mining,EMM-2016}, to use information theoretic measures~\cite{EMM-2016}, or to utilize likelihood-based approaches~\cite{EMM-evaluation}.
However, the respective functions are designed to quantify the exceptionality of a subgroup's induced model in a single dataset. 
Thus, as for the size component, we obtain an overall score for the exceptionality of the model in both datasets by aggregating the two scores for each dataset. Again, we suggest using the minimum of the two exceptionality scores to require both models to be unusual. However, alternatives such as the mean value or the geometric mean are viable alternatives.

\noindent\textbf{Similarity measures.}
Similarity measures $\phi_{\text{sim}}$ between two models that are derived by fitting the subgroup models in two different datasets have to our knowledge not yet been discussed in literature.
In principle, similarity measures can be constructed based on any measure of exceptionality $\text{ex}(D, \Omega)$ that quantifies how exceptional a model fitted to subgroup $D$ is with respect to a model fitted to the entire dataset $\Omega$. 
Such exceptionality measures are readily available for many exceptional model classes from literature.
Given such a measure, two models can be considered as similar if one is not exceptional with respect to the other.
However, such measures are often not symmetric in their arguments.
Since the position of the two models derived from the two subgroups should be considered as exchangeable in our setting, we have to adapt non-symmetric measures to use them as a similarity function $\phi_{\text{sim}}$.
In general, we distinguish between \emph{direct comparison approaches}, i.e., measures that compare the models obtained for subgroups from either side directly and \emph{common model approaches}, in which we compute a joint model on the union of both subgroups to compare against.

For \emph{direct comparison approaches}, given a symmetric measure of exceptionality that directly compares two models we suggest to construct a similarity measure $\text{sim}_{\text{d}}$  by computing 
\begin{align*}
\text{sim}_{\text{d} }(L, R) = \left(ex(L, R) + \epsilon  \right)^{-1}
\end{align*}
We hereby used $L/R$ to indicate subgroups coming from different datasets. The addition of $\epsilon$ avoids a division by zero.
When using an asymmetric measure of exceptionality for \emph{direct comparison approaches}, we propose to aggregate the exceptionality with interchanged arguments prior to inverting:
\begin{align*}
    sim_{\text {ex}}(L, R) = \left(\text{agg}\left(\text{ex}(L, R),\, \text{ex}(R, L) \right)+\epsilon \right)^{-1},
\end{align*}
where $agg$ is an aggregation function such as the minimum or (geometric) mean.
Often, exceptionality measures $ex$ are parameter-based, i.e., they compute a distance function such as Manhattan-distance or Euclidean-distance on the parameter values of the two models.
As another direct comparison measure, we propose the \emph{Crossed Likelihood similarity}, which we
designed based on model-based Subgroup Discovery~\cite{EMM-evaluation}. It aggregates the average likelihood of points from either side being generated by the model from the other side:
\begin{align*}
sim_{likely}(L, R)=\text{agg}\left(\frac{1}{|L|}\sum_{l \in L} p(l| \Theta_R),\, \frac{1}{|R|}\sum_{r \in R} p(r| \Theta_L)\right)
\end{align*}
where $p(x| \Theta_Y)$ is the likelihood of instance $x$ given  model parameters $\Theta_Y$.

In contrast to direct comparison approaches, \emph{common model approaches} measure the distance from a model which is fitted to the union of both subgroups.
We thereby enable the use of exceptionality measures which assume subset relationships in their arguments, i.e., they compare a subset of the data with the entire data.
To adapt such functions to similarity measures, we first compute the model parameters that are induced by the joint set of instances contained in the subgroups for the left and right datasets respectively, and then compute the corresponding exceptionality score before inversion.
\begin{align*}
    sim_{\text {ex}} = \left(\text{agg}\left(ex(L, L \cup R),\, ex(R, L \cup R) \right)+\epsilon \right)^{-1}
\end{align*}
Note that even though data instances from the left and the right datasets do not share all attributes, they do share the model attributes, allowing to compute the respective model for the union.
We suggest a common model similarity measure specifically for regression-like model classes, which is based on the popular Cooks distance, cf. \cite{EMM-slopes}. 
Cook's distance (roughly speaking) measures the influence of a set of points in linear regression.
Our similarity measure compares the models from either side against a model fitted to the union of points from both sides.
\begin{align*}
sim_{\text{Cooks}}(L,R) = \left( \text{agg}\left( ex_{C_{\Omega_L}}(L, L\cup R), ex_{C_{\Omega_R}}(R, L\cup R)\right) +\epsilon\right)^{-1}
\end{align*}
To measure all deviations with the same constant scale, we use in our experiments a fixed covariance matrix obtained from the entire data-set (indicated through the $\Omega_X$ subscript).

\subsection{Algorithms}
For detecting interesting patterns with Redescription Model Mining, we need to search in the product space of two description languages rather than in a single description language. For each pair of descriptions, we would then need to fit the respective two models, and compute the subgroups sizes, model exceptionalities, and model similarity to calculate the interestingness score.

The naive exhaustive way of approaching such a scenario is to enumerate all candidate patterns (pairs of descriptions) from both dataset.
As the pairs of candidates come from the product of two potentially large description languages, this approach quickly becomes infeasible.

To address this, we present a \emph{mine-and-pair} mining approach inspired by a corresponding algorithm from Redescription Mining~\cite{RDM-Gallo}.
The key idea of the mine-and-pair approach is that we first identify  promising candidate subgroups for each dataset separately by using a heuristic to reduce the number of potential candidates prior to matching candidates from both sides.
Thus, for Redescription Model Mining, we first perform traditional top-$k$ Exceptional Model Mining on both datasets (``mine-step''). 
For each result description, we cache relevant properties such as induced model parameters and subgroup representations to accelerate the subsequent pair-step.
In the pair-step, we then calculate the interestingness scores for all combinations of all subgroup candidates (i.e., the full cross product) in both datasets.
Note that the exhaustive approach can be considered as a special case of the mine-and-pair approach that is obtained by setting top-$k$ in the mine-step to a maximum value. 
By choosing lower values for the initial mine-step, the overall runtime can be reduced at the cost of potentially missing result patterns with high quality.
For common model approaches, we can achieve additional substantial speed-ups for many model classes by precomputing valuation bases, cf. \cite{EMM-trees}. These store sufficient statistics of the two disjoint sets and allow to compute the model parameters of the joint set in constant time.

As in other pattern mining tasks, a challenge that we encounter in Redescription Model Mining is redundancy in the result, i.e., several top subgroups are similar to each other in their descriptions or models.
To approach this issue, we can transfer different techniques for avoiding redundancy from the Exceptional Model Mining literature. For example, we could employ a specialized beam-search strategy in the mine-step to obtain a diverse set of patterns~\cite{van2012diverse} or adapt generalization-aware interestingness measures~\cite{lemmerich2013} for quantifying exceptionalities. For our experiments with real-world data, we relied on a basic technique that reduces redundancy, i.e., we employed a simple filter that keeps only the highest quality occurrence of the same subgroup description per dataset.

\section{Evaluation}
\label{sec:evaluation}
We evaluate\footnote{Code is available at \url{https://github.com/Feelx234/pyRDMM}} our approach on artificial as well as real-world data. While our approach can easily be applied to other model classes from the EMM literature, here, we focus on the correlation and regression model classes.
Experiments on artificially generated datasets allow for a quantitative evaluation because we have knowledge on the implanted patterns.
Thereafter, we demonstrate our method on two real-world datasets to illustrate the applicability of Redescription Model Mining for two specific cases:
First, we investigate house pricing patterns to illustrate a case where datasets on the same topic are collected independently of each other.
Second, we use two iterations of the European social survey to illustrate a case where data collection evolved over time and where we are able to associate questions that never appeared in the same survey.

\subsection{Experiments with Synthetic Data}
Experiments with synthetic data allow us to evaluate quantitatively to which degree Redescription Model Mining is able to recover inserted patterns in the presence of noise.
This is particularly important as there are no datasets with ground truth correspondences available.
We outline the generation of synthetic datasets before we describe the results of the conducted experiments.

\subsubsection{Generating artificial datasets}
\label{generate}

For conciseness, we only give a simplified outline of the data generation here and refer to the commented generating code in the codebase for details. 
Overall, we generate two datasets with eleven sets of instances each.
These sets are associated with individual model parameters for the respective model class (e.g., correlation or regression).
The eleven sets of parameters are the same for both datasets and correspond to one background model $\Theta_\text{regular}$ as well as ten sets of parameters $\Theta^i_\text{exceptional}$ of exceptional models.
For generating data for these models, additional "realization parameters" may be required, e.g., scale and variance, which are chosen appropriately.
Based on this, data is independently generated for each set of instances and each dataset.
For the background model, ten times more instances are generated than for the exceptional ones.
Then, we generate 20 binary describing attributes for each dataset such that each exceptional set of instances can be identified by exactly one conjunction of these attributes. 
We further add ten more binary describing attributes corresponding to random noise with different probabilities.
By employing Redescription Model Mining, we aim to recover the description pairs that correspond to the same exceptional model parameters.

\subsubsection{Regression model}
This section describes the experiments that were conducted on synthetic data using regression models.

\noindent\textbf{Setup. } 
For regression models, we use slope and intercept as model parameter sets $\Theta$ and generate data as described above.
We tested three different exceptionality measures and corresponding similarity measures:
Crossed likelihood distance, \emph{common model} Cook's distance, and
\emph{direct} parameter difference, i.e., the sum of absolute differences in the fitted model parameters.
To achieve robust results, we repeated the data generation and mining process ten times.

\begin{figure}[ht]
\vspace{-5mm}\includegraphics[width=0.99\columnwidth]{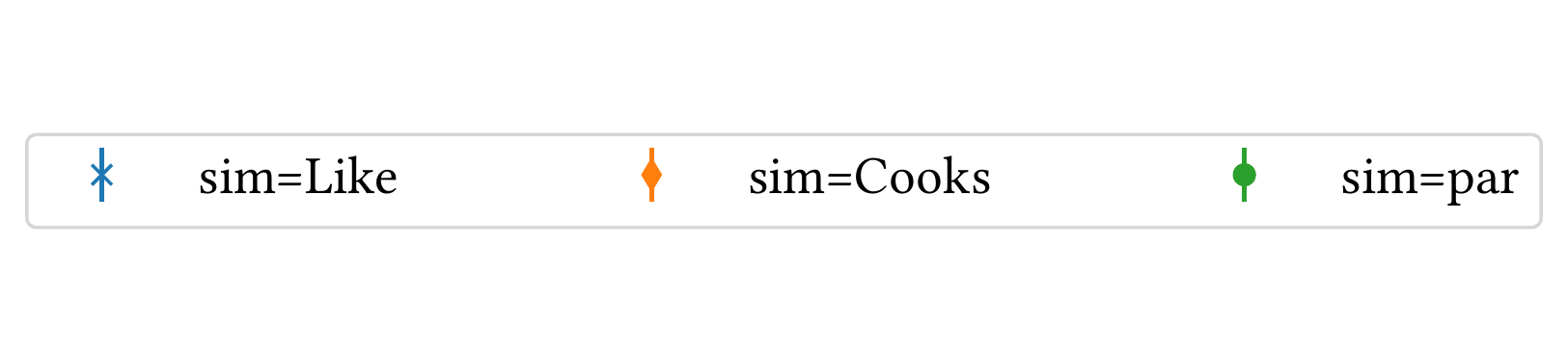}\vspace{-4mm}
		\begin{subfigure}[b]{0.528161994\columnwidth}
			\includegraphics[width=\textwidth]{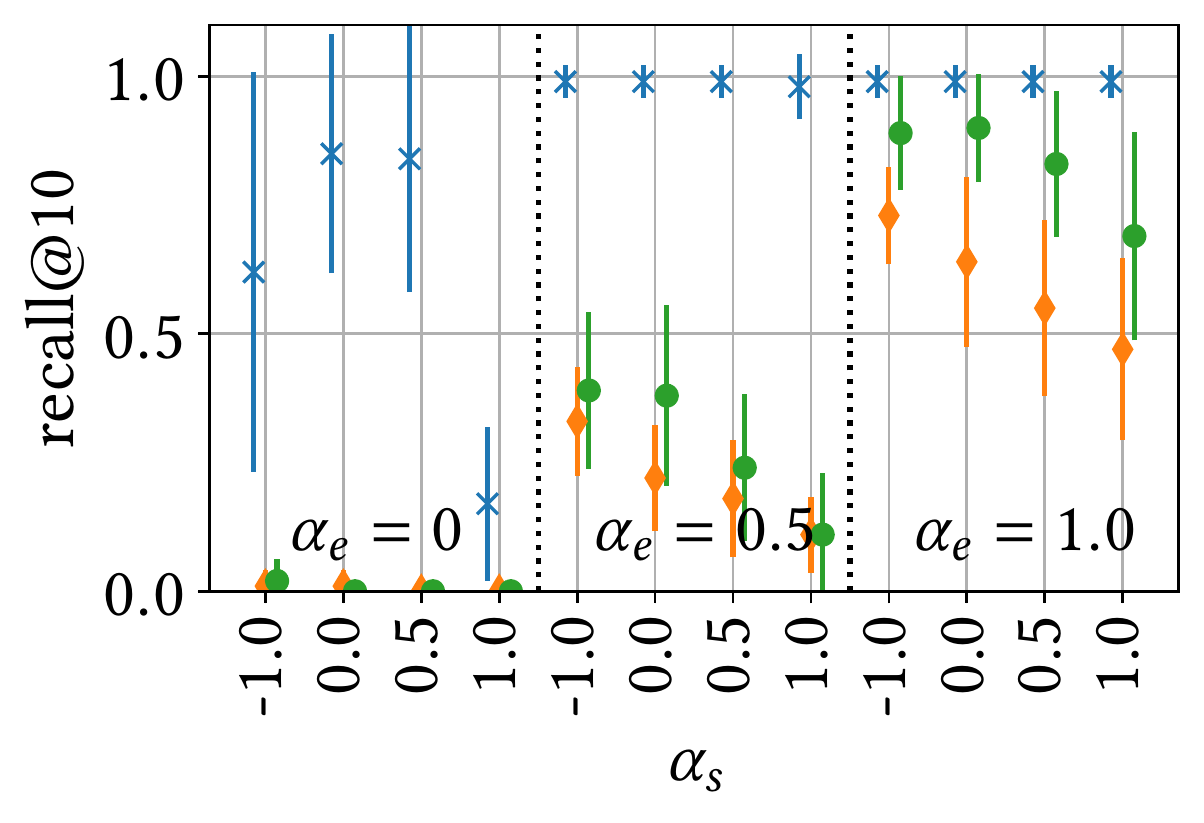}\vspace{-2mm}%
			\caption{Exhaustive, ex=Like}\vspace{2mm}%
		\end{subfigure}
		\begin{subfigure}[b]{0.451838006\columnwidth}
			\includegraphics[width=\textwidth]{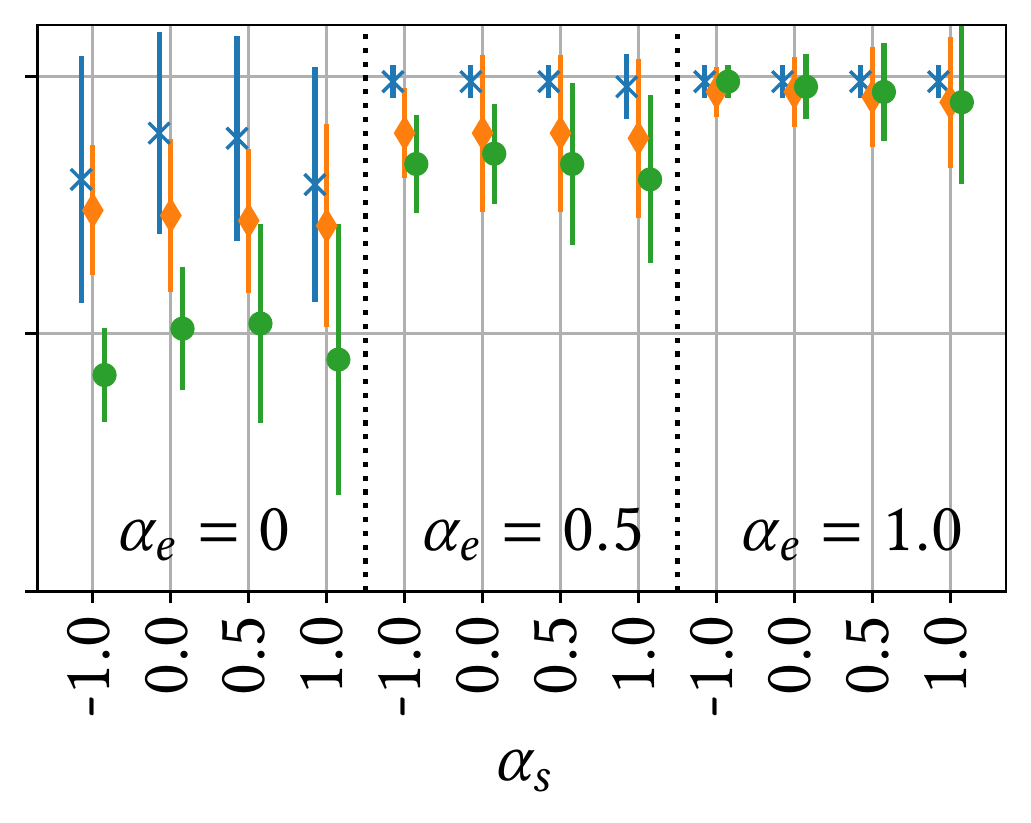}\vspace{-2mm}%
			\caption{Mine-pair, ex=Like}\vspace{2mm}%
		\end{subfigure}
		\begin{subfigure}[b]{0.528161994\columnwidth}
			\includegraphics[width=\textwidth]{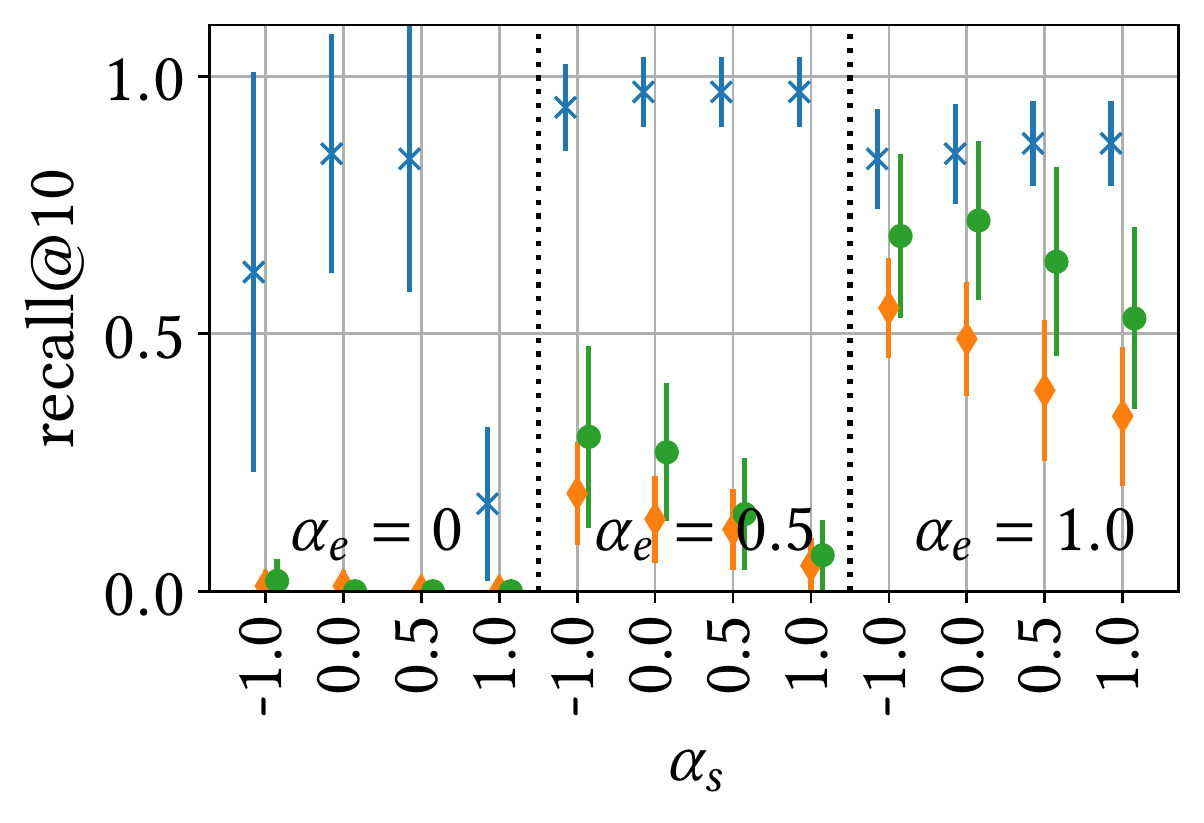}\vspace{-2mm}%
			\caption{Exhaustive, ex=Cooks}\vspace{2mm}
		\end{subfigure}
		\begin{subfigure}[b]{0.451838006\columnwidth}
			\includegraphics[width=\textwidth]{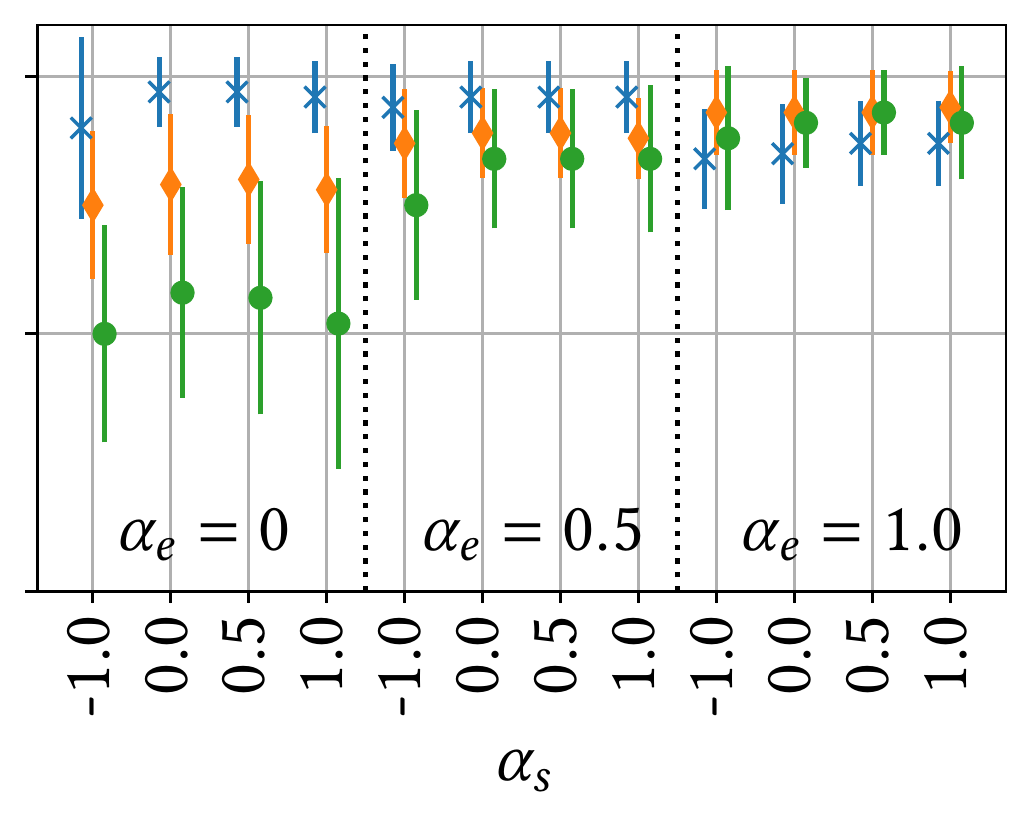}\vspace{-2mm}%
			\caption{Mine-pair, ex=Cooks}\vspace{2mm}%
		\end{subfigure}
		\begin{subfigure}[b]{0.528161994\columnwidth}
			\includegraphics[width=\textwidth]{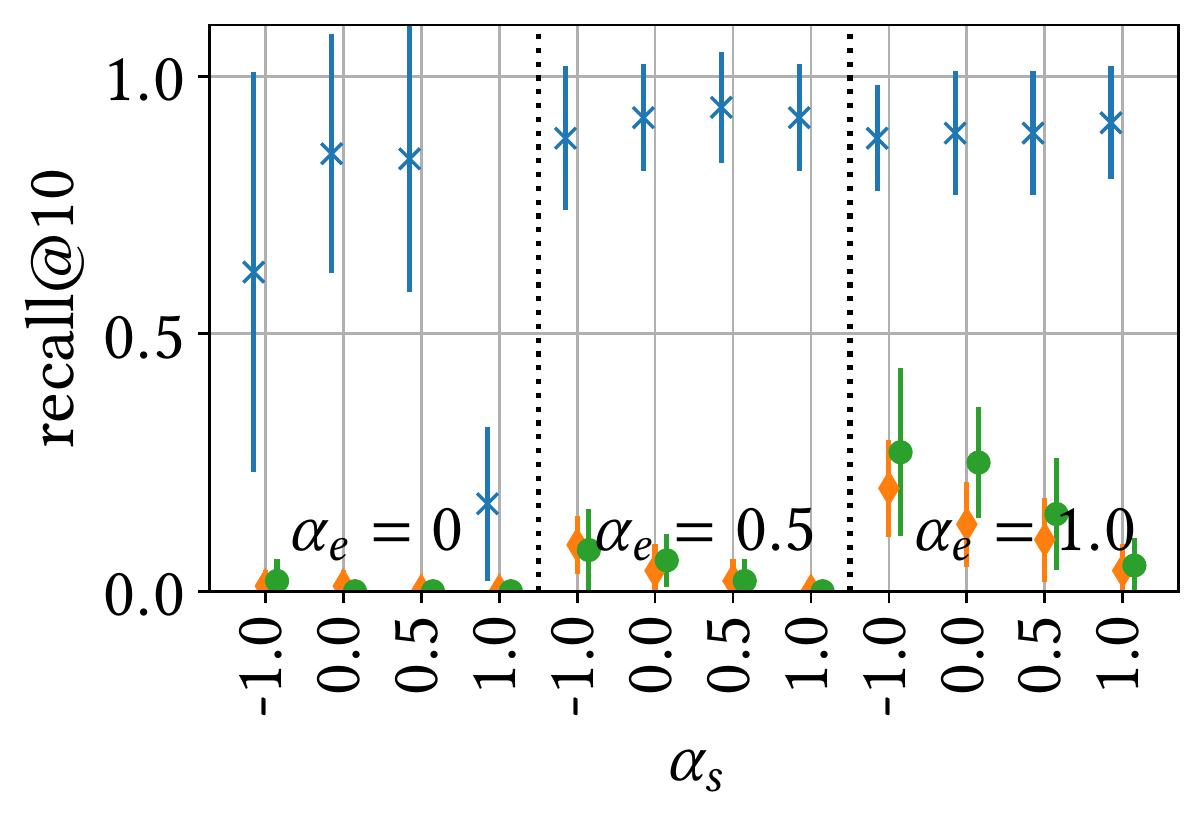}\vspace{-2mm}%
			\caption{Exhaustive, ex=par}%
		\end{subfigure}
		\begin{subfigure}[b]{0.451838006\columnwidth}
			\includegraphics[width=\textwidth]{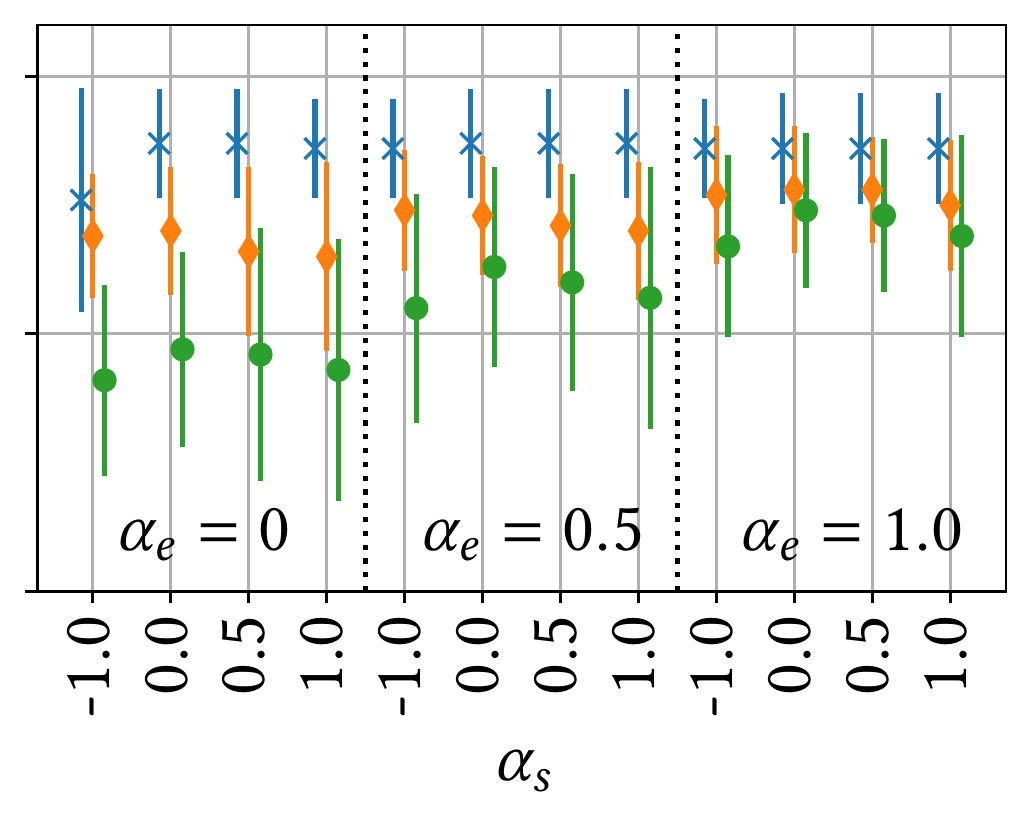}\vspace{-2mm}%
			\caption{Mine-pair, ex=par}%
		\end{subfigure}
		\caption{Synthetic experiments for the linear regression model class. The y-axis denotes recall@10, i.e., how many of the implanted patterns are recovered. The left column shows results using the exhaustive search algorithm while the right column shows results for the mine-and-pair approach. In different rows we varied the exceptionality measure. Each plot is separated into three horizontal sections which correspond to different values of the exceptionality scaling parameter $\alpha_e$. Within those sections, we vary the size scaling $\alpha_s$. Displayed are mean and standard deviation obtained from runs on ten synthetic dataset pairs.
		Overall, the mine-and-pair approach outperforms the exhaustive approach for similar parameter settings. Also increasing the exceptionality scaling increases recall.}
	\label{synthreg}	\vspace{-4mm}
\end{figure}

\noindent\textbf{Results. }
Regarding the runtimes of the exhaustive and mine-and-pair mining algorithms, we observe that mine-and-pair is by multiple orders of magnitude faster with a total runtime of less than 15 minutes on a standard desktop machine compared to about 20 hours for the exhaustive approach.
To evaluate the performance of our approach, we measure the recall@10 of the returned patterns, i.e., how many of the implanted patterns are returned in the top 10 results.
In Fig.~\ref{synthreg}, we compare combinations of algorithms, interestingness measures, and scaling parameters $\alpha_s$ (size) and $\alpha_e$ (exceptionality). The left column shows results for the exhaustive approach while the right column for the mine-and-pair approach. Per row, we display the results for one exceptionality measure. Within each plot, there are three sections corresponding to increasing values of the exceptionality scaling $\alpha_e$. Within each section, we show results for different values of the size scaling exponent $\alpha_s$. For better readability, we have slightly offset observations made for the same $\alpha_s$ value.

When comparing similarity measures for the same algorithm and exceptionality, the similarity measure based on likelihood (sim=Like) in almost all cases outperforms the alternatives.
This may be explained by the fact that the likelihood-based similarity takes into account the models' goodness of fit.
Using the exhaustive approach, we can see that results for $\alpha_e = 0$ (i.e., disabling the exceptionality measure completely as shown in the left section of each plot) yields generally bad results for the other two similarity measures.
When increasing $\alpha_e$ (with other parameters constant), we can observe that the recall increases on average. 
As argued in Section~\ref{sec:interestingness}, this observation is expected as the exceptionality is needed to avoid random findings.
This effect is less pronounced for the mine-and-pair approach even though still present.
This is due mine-and-pair's restriction to subgroups with high exceptionality in the mine-step thus ruling out model pairs that are extremely similar but not exceptional in the pair-step.
In contrast, such model pairs are considered in the exhaustive approach.

The effect that an increase in $\alpha_s$ with otherwise fixed parameters leads to a decrease in recall for the exhaustive approach but not for mine-and-pair is less well understood.
This may be explained by the effect that patterns that are extremely similar across datasets but not exceptional also tend to be large.
Such patterns are already ruled out in the mine-and-pair approach thus mitigating this effect.

\subsubsection{Correlation}
This section outlines the experiments conducted for the correlation model class.

\noindent\textbf{Setup. }
For our synthetic experiments with the correlation model class, we generate 5 by 5 covariance matrices as model parameters $\Theta$. We further sample model attribute values from a multidimensional Gaussian distribution with that covariance matrix.
Otherwise, we use the procedure outlined in Section~\ref{generate}. For further details, we refer to the supplementary material.
As for regression, we repeated the data generation and mining process ten times.

\begin{figure}[ht]
		\begin{subfigure}[b]{0.528161994\columnwidth}
			\includegraphics[width=\textwidth]{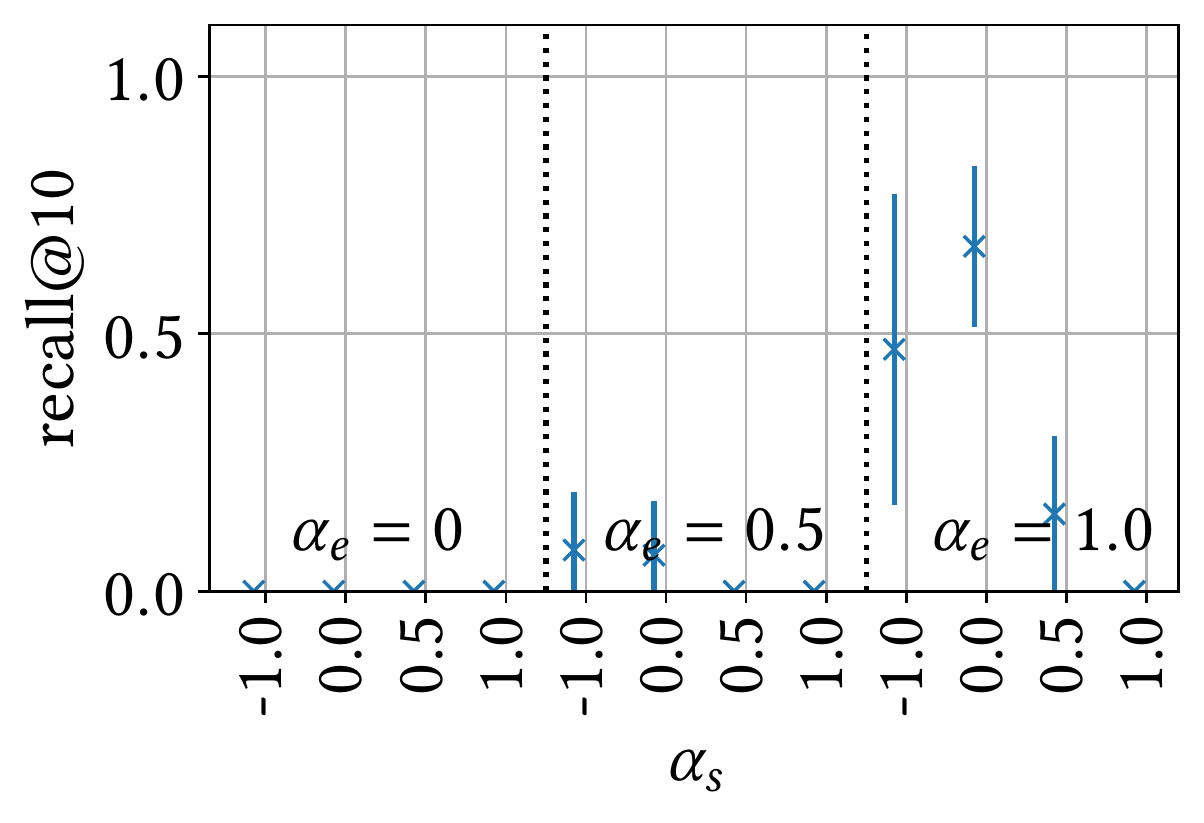}
			\caption{Exhaustive approach}
		\end{subfigure}
		\begin{subfigure}[b]{0.451838006\columnwidth}
			\includegraphics[width=\textwidth]{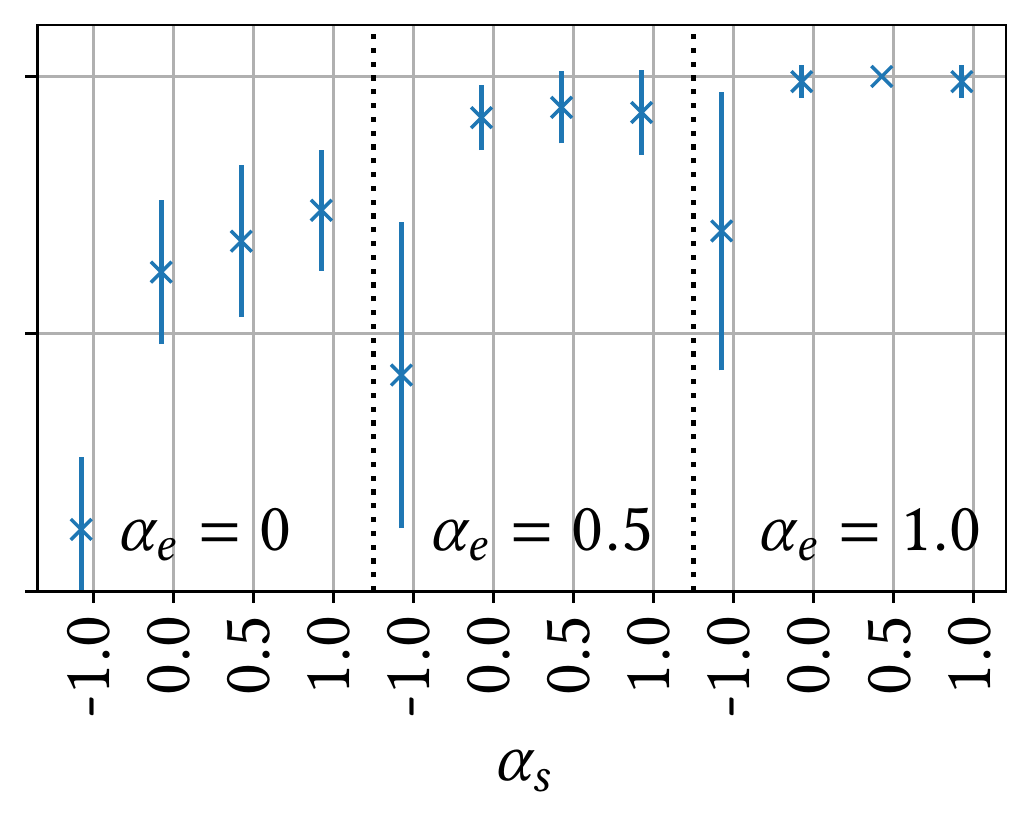}
			\caption{Mine-pair approach}
		\end{subfigure}
		\caption{Synthetic experiments for the correlation model class. The y-axis denotes recall@10, i.e., how many of the implanted patterns are revered. We use the 1-norm of the matrix difference for both similarity and exceptionality. The results are averaged over ten dataset pairs. Visualized are the mean and standard deviation. Increasing focus on exceptionality ($\alpha_e$) improves the recall for both the exhaustive and mine-and-pair approach. Also the mine-and-pair approach outperforms the exhaustive approach for the same parameters. This shows that correlation patterns can be successfully extracted from pairs of datasets when choosing the right parameters, thus, illustrating the flexibility of our approach across model classes.}\vspace{-5mm}
\end{figure}

\noindent\textbf{Results. }
For the correlation model class, we limited the experiments to a single combination of exceptionality and similarity measure.
For similarity and exceptionality, we choose a direct comparison 1-norm, i.e., we take the 1-norm of the difference in correlation matrices.
Similar to the linear regression experiments, increased values of $\alpha_e$ lead to better results for both the exhaustive and mine-and-pair approach with the exhaustive approach having worse results overall.
Analogously, a decrease in $\alpha_s$ leads to better performance for the exhaustive approach, while this is not true for mine-and-pair.
Similar explanations as in the exhaustive case hold.

\subsubsection{Summary}
The evaluation on synthetic datasets shows that, given the right parameters, both algorithmic approaches can reliably recover the implanted patterns.
The experiments also reveal that using the exceptionality term is essential to achieve good performance for both approaches.
Overall, the mine-and-pair approach mostly outperforms the exhaustive approach due to its implicit minimal exceptionality threshold, i.e., we consider only the top $k$ most exceptional subgroups during the mine step. 
At first glance, it is counterintuitive that the heuristic mine-and-pair outperforms the exhaustive approach. However, the exhaustive approach only returns improved results with respect to scores of the employed interestingness measures, not the recall of actual non-random patterns. If the measures themselves are misconfigured, higher scoring patterns in the results do not lead to higher recalls with respect to the generated patterns. As highlighted by our experiments, finding a ``correct'' configuration can be challenging in practice. We observe that in many settings, the mine-and-pair approach will give better results in terms of recall when the interestingness measure is misaligned since the initial mine step serves as a filter.

\subsection{Experiments with Real-world Data}
In this section, we demonstrate Redescription Model Mining applied to two real-world datasets.
The goal is to demonstrate that the proposed techniques can be applied to real-world data rather than to gain actionable insight into the respective domains. 
The investigated dataset combinations are from different themes and with different models.
Firstly, we mine corresponding exceptional regression models in datasets on house prices crawled from the web.
Thereafter, we find corresponding exceptional correlation models in the European social survey. Parameters of the interestingness measures have been determined through an interactive refinement process.

\subsubsection{Housing}
Studying house prices is a common example application in Exceptional Model Mining~\cite{duivesteijn2012different,EMM-2016}.
We used two datasets that have rich sets of descriptive attributes and a large number of instances.
One dataset is on houses in the city of Melbourne, Australia, sold from Domain.com.au\footnote{ \url{https://www.kaggle.com/dansbecker/melbourne-housing-snapshot}}.
The second dataset is on housing data for the city of Beijing, China, sold via bj.lianjia.com\footnote{ \url{https://www.kaggle.com/ruiqurm/lianjia}}.
We apply minimal preprocessing and removal of outliers. For more details, please refer to the supplementary material.

\begin{figure}[ht]
	\begin{center}
		\begin{subfigure}[b]{0.47\columnwidth}
			\includegraphics[width=\textwidth]{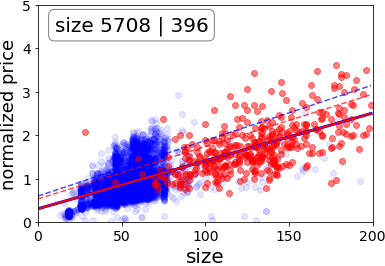}
			\caption{bedroom=1 $\wedge$  district=6 $\leftrightarrow$\\ type=`t' $\wedge$  yearBuilt$\geq$2000.0}
			\label{fig:results-a}
		\end{subfigure}\hspace{2mm}
		\begin{subfigure}[b]{0.47\columnwidth}
			\includegraphics[width=\textwidth]{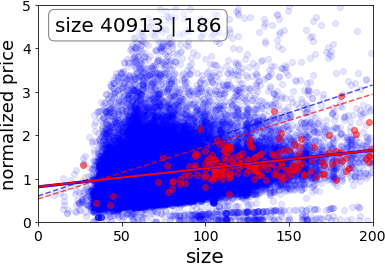}
			\caption{structure=2 $\wedge$  renovation=1 $\leftrightarrow$ area=`Brimbank'}
			\label{fig:results-b}	
		\end{subfigure}
		\begin{subfigure}[b]{0.47\columnwidth}
			\includegraphics[width=\textwidth]{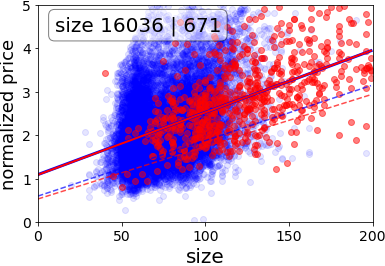}
			\caption{DOM:$[15.0,48.0[$ $\wedge$  bedroom=2 $\leftrightarrow$ method=`S' $\wedge$  yearBuilt $<$ 1930}
			\label{fig:results-c}
		\end{subfigure}\hspace{2mm}
		\begin{subfigure}[b]{0.47\columnwidth}
			\includegraphics[width=\textwidth]{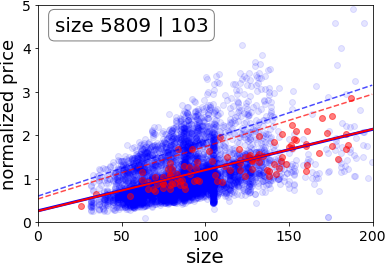}
			\caption{district=11 $\wedge$  elevator=False $\leftrightarrow$ area=`Mariby' $\wedge$  yearBuilt$\geq$ 2000}
			\label{fig:results-d}
		\end{subfigure}
		\begin{subfigure}[b]{0.47\columnwidth}
			\includegraphics[width=\textwidth]{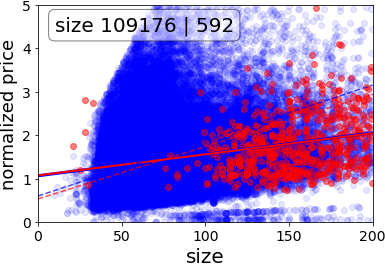}
			\caption{structure=2 $\wedge$ kitchen = 1 $\leftrightarrow$ \\ bathroom=2  $\wedge$\\  distance $\geq$ 13.80}
			\label{fig:results-e}
		\end{subfigure}\hspace{2mm}
		\begin{subfigure}[b]{0.47\columnwidth}
			\includegraphics[width=\textwidth]{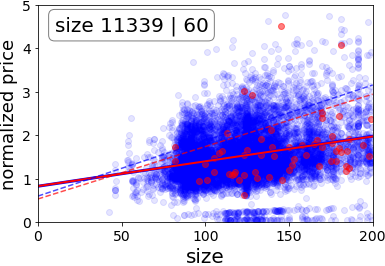}
			\caption{district=6 $\wedge$  livingroom=2: $[3,9[$ $\leftrightarrow$ car$\geq$4 $\wedge$  yearBuilt: $[1970, 2000[$}
			\label{fig:results-f}
		\end{subfigure}
	\end{center}	\vspace{-2mm}
	\caption{Results of Redescription Model Mining on the housing datasets (linear regression model class). Visualized are the top findings according to a subjectively refined interestingness measure. The exceptionality measure is likelihood gain, the similarity measure is common model Cooks similarity. Blue/ red color indicates data from Beijing/ Melbourne. The dashed lines are fitted to the entire dataset while the solid lines are fitted to the respective subgroups. Instances in the subgroups are visualized as transparent circles. Redescription Model Mining identifies corresponding descriptions across Beijing and Melbourne with regard to the relation between the price and size of sold houses.}
	\label{fig:housing}
	\vspace{-5mm}
\end{figure}

The top 6 discovered patterns for common model cooks similarity and likelihood gain as exceptionality measure are visualized in Fig.~\ref{fig:housing}. Information on the Beijing/Melbourne dataset is coloured in blue/red respectively. Models fitted to the entire datasets/ to subgroups are visualized through dashed/solid lines. 
For all patterns, the solid lines overlay each other almost perfectly, thus they are not easy to distinguish. The subsets that correspond to those models are visualized as blue and red dots respectively.

The patterns exhibit a large variety in the found descriptions and models.
We identified some patterns with similar slope as the overall dataset but lower intercept (\ref{fig:results-a}, \ref{fig:results-d}) as well as higher intercept (\ref{fig:results-c}).
Comparing these two patterns, both descriptions based on data from Melbourne contain the yearBuild attribute with the earliest yearBuild interval corresponding to higher-price houses and the latest yearBuild interval for the on-average lower-price houses.
Both Beijing-based descriptions contain the bedroom attribute with bedrooms=2 corresponding to on-average more expensive houses and bedrooms=1 for lower priced houses.
We also observe patterns with significantly lower slopes, e.g., Fig.~\ref{fig:results-b}. 
Looking at the descriptions, we see that these houses are structure=2 (mixed) and renovation=1 (1 meaning 'other', probably the worst) on the Beijing side and Area='Brimbank' on the Melbourne side.
It seems that Brimbank is one of the less wealthy areas within the Melbourne Metropolitan region.
Thus, this correspondence also seems plausible.

\subsubsection{European social survey}
The European Social Survey (ESS) is conducted across Europe every two years.
The survey contains a set of questions that are shared across rounds while other questions change every round.
Redescription Model Mining allows for finding potential associations between responses to questions changing across years via modelling shared questions.
We demonstrate this by relating questions from Round 8~\cite{ESS8} to Round 9~\cite{ESS9}.

\begin{figure}[ht]
	\begin{center}
	\begin{subfigure}[b]{0.6\columnwidth}
			\includegraphics[width=\textwidth]{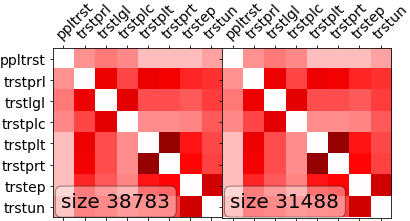}
			\caption{Complete Dataset}
		\end{subfigure}
		\begin{subfigure}[b]{0.47\columnwidth}
			\includegraphics[width=\textwidth]{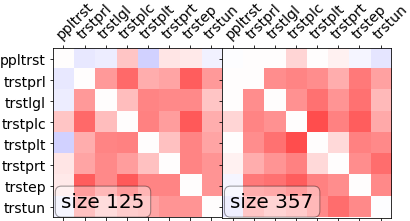}
			\caption{\smaller{elgnuc=`A small amount' $\wedge$  rdcenr = `Never' $\leftrightarrow$\\ occinfr =`Low, extremely unfair' $\wedge$  sofrprv =`Neit. agree nor disagree'}}
			\label{fig:results-3-1-0}
		\end{subfigure}\hspace{2mm}
		\begin{subfigure}[b]{0.47\columnwidth}
			\includegraphics[width=\textwidth]{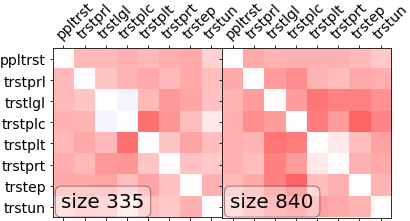}
			\caption{\smaller{rdcenr='Hardly ever'$\wedge$  wrntdis=`Not at all worried' $\leftrightarrow$\\ evmar=`Yes' $\wedge$  wltdffr=`Small, extremely unfair'}}
	
		\end{subfigure}
		\begin{subfigure}[b]{0.47\columnwidth}
			\includegraphics[width=\textwidth]{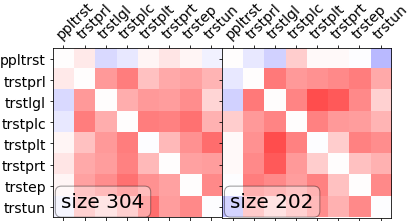}
			\caption{\smaller{mnrgtjb=`Agree' $\wedge$  wrdpfos=`Not at all worried' $\leftrightarrow$\\  gvintcz=`Not at all' $\wedge$  ifredu=4}}
		\end{subfigure}\hspace{2mm}
		\begin{subfigure}[b]{0.47\columnwidth}
			\includegraphics[width=\textwidth]{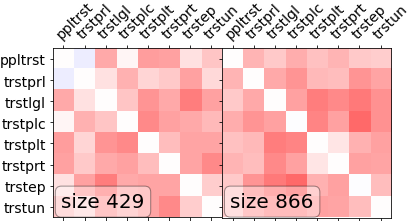}
			\caption{\smaller{ccgdbd=6 $\wedge$  inctxff=`Somewhat in favour' $\leftrightarrow$ bthcld=`Yes' $\wedge$  wltdffr=`Small, extremely unfair'}}
		\end{subfigure}
		\begin{subfigure}[b]{0.47\columnwidth}
			\includegraphics[width=\textwidth]{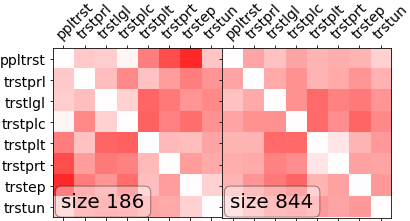}
			\caption{\smaller{clmchng=`Definitely not changing' $\wedge$  gvrfgap=`Agree' $\leftrightarrow$\\  btminfr=`Low, extremely unfair' $\wedge$  wltdffr=`Small, extremely unfair'}}
		\end{subfigure}\hspace{2mm}
		\begin{subfigure}[b]{0.47\columnwidth}
			\includegraphics[width=\textwidth]{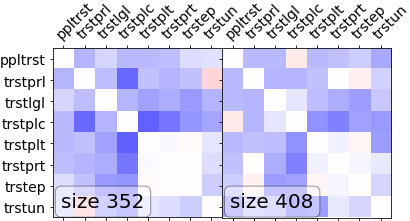}
			\caption{\smaller{gvsrdcc=1 $\wedge$  wrdpimp=`Very worried' $\leftrightarrow$\\ plnftr=3 $\wedge$  wltdffr=`Large, extremely unfair'}}
		\end{subfigure}

	\end{center}
	\caption{Results of Redescription Model Mining for the European social survey (ESS) datasets (correlation model class). We show heatmaps of correlation matrices for the entire dataset and top 6 result patterns. We subtracted the average of dataset matrices from the pattern matrices. The left/ right matrices in each subplot show the subgroup correlation from the 8th/ 9th iteration of the ESS. All matrices use the same scale for colours. The descriptions are presented in the figure captions. These uncovered patterns relate attributes which never appeared in the same iteration of ESS based on models fitted on attributes which do appear in both iterations.}
	\label{fig:ESS}	\vspace{-5mm}
\end{figure}

We use a correlation model on the attributes that assess participants' trust in specific (political) organs.
The describing attributes for Round 8 include questions about energy usage, climate change, and refugees.
For Round 9, describing attributes are taken from questions about justice and fairness.
Similar to the synthetic experiments, we use the 1-norm of covariance matrices for both exceptionality and similarity.
Results for the size scaling $\alpha_s=0.1$ and the exceptionality scaling of $\alpha_e=0.5$ are visualized in Fig.~\ref{fig:ESS}.

To give a better impression of what patterns are uncovered, we describe the meaning of the correspondence from Fig.~\ref{fig:results-3-1-0} as follows:
People, who say that a small amount of electricity should be generated from nuclear power and which never do things to reduce energy usage, show a similar exceptional correlation in their trust in political systems as people who think that it is extremely unfair how much net [pay/pensions/social benefits] other people of similar occupation receive in comparison and that they neither agree or disagree with whether society should take care of the poor regardless of what they give back.
As our method merely unveils candidates for further inspection, we leave more fundamental investigations and assessments of these patterns to field experts.

From a methodological perspective, the overall uncovered correspondences yield patterns where the correlation matrices are significantly different from the dataset correlation matrices.
This lends support to our choice of exceptionality and similarity measures, indicating the expressiveness and flexibility of our approach.

\clearpage
\section{Limitations}
\label{sec:discussion}

Redescription Model Mining is a novel pattern mining task that allows to identify correspondences across two disjoint datasets that share no (or few) instances and only a part of their attributes.
While we think that this approach can help to generate knowledge by connecting datasets within a domain in a completely novel way, researchers and practitioners should be aware of its limitations.
First and foremost, we emphasize that the presented approach only generates candidate correspondences which require further inspection whether they really unveil a common underlying phenomenon.
From the experiments on real-world datasets, it is evident that our approach ---as most pattern mining approaches--- can substantially benefit from domain experts to assess the quality of the findings and select the most promising candidates for detailed investigation.

In that direction, we assume an interesting relationship with respect to the complexity of the chosen model classes: if we find close similarities for a very complex model class with many parameters, a common underlying phenomenon seems intuitively to be more likely compared to similarities in simple model classes.

From a practical perspective, Redescription Model Mining is subject to well-known challenges from the pattern mining literature. This includes scalability of the mining approaches, reducing redundancy in the result set, and facilitating interactive approaches by providing a flexible and tunable framework. While this paper presents initial approaches to tackle these problems, we see our paper as a first step towards more in-depth studies in this novel field of research.

\section{Conclusions}
This work introduced Redescription Model Mining, a novel pattern mining approach to establish relationships across datasets, which share a small set of key domain-relevant attributes but no instances.
By combining ideas and techniques from Exceptional Model Mining and Redescription Mining, Redescription Model Mining finds corresponding pairs of descriptions which induce similar and exceptional models.
We introduced a framework for interestingness measures, which allows for weighting model similarity, exceptionality, and size against each other to facilitate the discovery of interesting correspondences across datasets and proposed different options for adapting measures to capture those individual components. 
Furthermore, we developed a mine-and-pair approach that allows for efficiently mining such patterns.
We demonstrated the potential of our approach on several synthetic and real-world datasets.
Overall, redescription model mining unveils candidates for relationships between attributes which are not observed together in the same dataset or for the same instances. This allows for connecting information from previously incompatible data sources. 
To the authors' knowledge, this work presents the first pattern mining approach able to utilize such weakly connected datasets.

Our work enables multiple directions for future research.
The development of efficient search algorithms, reducing redundancy and providing further guidance for selecting the interestingness measures would facilitate practical applications. In that direction, full case studies involving domain experts could further demonstrate the usefulness of the presented approach, but could also reveal new practical challenges.


\begin{acks}
We thank P. Carloni for his support in realizing this research project. 
\end{acks}

\bibliographystyle{ACM-Reference-Format}
\bibliography{bib}



\end{document}